\documentclass[epj]{svjour}
\usepackage{graphics}
\begin{document}
\title{Direct Fragmentation of Quarkonia Including Fermi Motion Using Light-cone Wave Function}
\author{M.A. Gomshi Nobary\inst{1,2} \and B. Javadi\inst{1}
}%
%
%
\institute{Department of Physics, Faculty of Science, Razi
University, Kermanshah, Iran; \and The Center for Theoretical
Physics and Mathematics, A.E.O.I., Chamran Building, P.O. Box
11365-8486 Tehran, Iran.}
\date{Received: date / Revised version: date}
%
\abstract{
 We investigate the effect of Fermi motion on the direct
 fragmentation of the $J/\psi$ and $\Upsilon$ states
 employing a light-cone wave function. Consistent with such a wave
 function we set up the kinematics of a heavy quark fragmenting
 into a quarkonia such that the Fermi motion of the constituents
 split into longitudinal as well as transverse direction and thus
 calculate the fragmentation functions for these states. In the framework of our investigation, we
 estimate that the fragmentation probabilities of $J/\psi$ and
 $\Upsilon$ may increase at least up to 14 percent when including this
 degree of freedom.
} 
\authorrunning {M.A. Gomshi Nobary and B. Javadi}
\titlerunning {Direct Fragmentation of Quarkonia...}
\maketitle
%
\authorrunning {M.A. Gomshi Nobary and B. Jvadi}
\titlerunning {The Role of Fermi Motion in Quarkonia Fragmentation}
\maketitle
\section{Introduction}
\label{intro}

Quarkonia production in high energy hadron collisions have shown
to be an important issue both in QCD and in collider physics.
Hadronic production of $J/\psi$ and $\Upsilon$ have received
attention since the discovery of charm and beauty quarks. The
production of these states have been studied in interesting models
[1]. It is also found that there is an order of magnitude
discrepancy between cross section data and theoretical
calculations in the case of $J/\psi$ at the Tevatron energies.
Then it became clear that the dominant mechanism at high energy
collisions are the color singlet [1,2] and the color octet [3]
mechanisms and more over that at the Tevatron energies the color
singlet contribution is too small [4,5]. In the color singlet
mechanism it is assumed that the $Q\bar Q$ pair is produced in
color singlet state at the production point whereas in the later
color octet case the $Q\bar Q$ pair is produced in the form of
color octet. Later on emission of a soft gluon takes the pair into
a color singlet state. To produce singlet or octet $Q\bar Q$ there
will of course be various channels out of which direct quark and
gluon fragmentation are thought to be the dominant ones. In each
case we need fragmentation functions to evaluate the probabilities
and cross sections.

The very accurate and successful calculation of the fragmentation
functions have recently been carried out in the limit of heavy
quark effective theory. It has been proved that the strong
interaction perturbation theory is suited to calculate the hard
part of the heavy meson fragmentation in the above limit [6] and
use a kind of wave function to introduce the soft smearing of the
bound state. These are then convoluted to give the fragmentation
function. This convolution is possible only in a certain QCD
scale. The Altarelli-Parisi evolution equation [7] takes this
function into a required scale. In the case of fragmentation of
heavy mesons, either a kind of delta function or the wave function
at the origin of the bound state have been employed which produces
a factor of meson decay constant, $f_M$, or the wave function at
the origin $|R(0)|$ and lets the constituents move almost together
in parallel ignoring their virtual motion. However the problem of
$J/\psi$ cross section at Tevatron mentioned earlier and also its
diffractive production [8] require a closer look at the process.

The common procedure for introducing the bound state effects, is
to use the wave function at the center of momentum frame in a
harmonic oscillator model. However the kinematics and
parameterization in such a case is not much clear. On the other
hand the procedure has successfully been studied in an infinite
momentum frame. And that Brodsky-Huang-Lepage (BHL) [9] have
proposed a boost free wave function for the heavy meson bound
state which fits in the above calculational scheme. This is a kind
of gaussian wave function with a parameter which adjusts its width
is referred to as the confinement parameter. The mass of the
constituents and their transverse momentum as well as the
longitudinal energy-momentum ratios appear in this function.

In this work we first describe the BHL wave function. Using it we
calculate the fragmentation function for the $J/\psi$ and
$\Upsilon$ states in both cases i.e. when the Fermi motion is
either ignored or not ignored. Change of the confinement parameter
provides a nice comparison of these two functions. It is revealed
that the inclusion of this degree of freedom may have a sizable
effect on fragmentation probabilities and hence on their cross
section measured in a particular machine aimed to produce such
states.

\section{Heavy meson wave function}
\label{sec:2} The light cone wave functions described in this
section have satisfactorily been applied for light meson states
such as $\pi$, $\rho$ and $K$ [10]. motivated by these studies we
will apply them to the case of $J/\psi$ and $\Upsilon$ states.

The mesonic wave function can be obtained by solving the
bound-state equation [11]
\begin{eqnarray}
(M^2-H_{LC})|{\cal{M}}\rangle=0,
\end{eqnarray}
where $H_{LC}$ is the hamiltonian in light-cone quantization and
$M$ is the mass of the meson $|\cal{M}\rangle$. In an attempt to
solve this equation one expands $|\cal{M}\rangle$ in the complete
set of Fock states

\begin{eqnarray}
|{\cal{M}}\rangle=\sum_{n,\lambda_i}\psi_n(x_i,{\bf
q}_{T_i},\lambda_i)|n;x_i,{\bf q}_{T_i},\lambda_i\rangle.
\end{eqnarray}
In the above relation $\psi_n(x_i,{\bf q}_{T_i},\lambda_i)$,
$(i=1,2,...,n)$ is the $n$-parton momentum-space amplitude defined
on the free quark and gluon Fock basis in the light-cone
formalism. The normalization condition is

\begin{eqnarray}
\sum_{n,\lambda_i} \int[dx][d^2{\bf q}_T]|\psi_n(x_i,{\bf
q}_{T_i},\lambda_i)|^2=1,
\end{eqnarray}
where
\begin{eqnarray}
[dx]\equiv \prod_{i=1}^n dx_i\delta\Bigl[1-\sum_{i=1}^n x_i\Bigr],
\end{eqnarray}
and
\begin{eqnarray}
[d^2 {\bf q}_T]\equiv \prod_{i=1}^n d^2{\bf q}_{T_i} 16\pi^3
\delta^2\Bigr[\sum_{i=1}^n {\bf q}_{T_i}\Bigl].
\end{eqnarray}
The sum is over all Fock states and helicities.

It is not easy to determine the form of the wave functions from
the above expansion. The prescription due to Brodsky-Huang-Lapage
(BHL) [9] provides a practical way by connecting the equal time
wave function in the rest frame and the light-cone wave function,
i.e. by equating the off-shell propagator
$\epsilon=M^2-\bigl(\sum_{i=1}^n q_i\bigr)^2$ in two frames:
\begin{eqnarray}
\epsilon&=&M^2-\bigl[\sum_{i=1}^np_i^\circ \bigr]^2\;\;[{\rm
c.m.}]\nonumber\\&&=M^2-\sum_{i=1}^n\bigl[({\bf
q}_{Ti}^2+m_i^2)/x_i\bigr]\;\;[\rm{L.C.}]
\end{eqnarray}
where the constraints $\sum_{i=1}^n {\bf p}_i=0$ and
$\sum_{i=1}^n{\bf q}_{Ti}=0$, $\sum_{i=1}^n x_i=1 $ apply for c.m.
and L.C. frames respectively. It is easily seen that for a two
particle system with equal mass $(p_1^\circ=p_2^\circ)$

\begin{eqnarray}
{\bf p}^2\longleftrightarrow {{{\bf q}_T^2+m^2}\over
{4x(1-x)}}-m^2,
\end{eqnarray}
where $p$ is the moment of the constituents in the center of
momentum frame. Therefore the wave functions at two frames may be
connected by

\begin{eqnarray}
\psi_{\rm c.m.}({\bf p}^2)\longleftrightarrow
\psi_{LC}\biggl[{{{\bf q}_T^2+m^2}\over {4x(1-x)}}-m^2\biggr],
\end{eqnarray}
Now if we consider the wave function at the rest frame as [12]

\begin{eqnarray}
\psi_{\rm c.m.}({\bf p}^2)=A_M\exp\biggl[
\frac{-p^2}{2\beta^2}\biggr],
\end{eqnarray}
where $\beta$ is the confinment parameter, then using the
connection (8) the $IMF$ wave function in the case of equal mass
constituents takes the following form

\begin{eqnarray}
\psi_M(x_i,{\bf q}_T)&=&A_M\exp
\Biggl\{-{1\over{8\beta^2}}\biggl[{m_T^2\over{x_1}}+
{m_T^2\over{x_2}} \biggr]\Biggl\}\nonumber\\
&&=A_M\exp\biggl[ \frac{m_T^2}{8\beta^2 x_1(1-x_1)} \biggr].
\end{eqnarray}
Note that in this formalism ${\bf q}_{T_1}=-{\bf q}_{T_2}$,
therefore $q_{T_1} ^2=q_{T_2} ^2 =q_T^2$. It is argued that this
form holds in the case of heavy meson states with equal and
unequal constituent quark and antiquark masses, i.e.

\begin{eqnarray}
\psi_M(x_i,{\bf k}_T)=A_M\exp
\Biggl\{-{1\over{8\beta^2}}\biggl[{{m_{1T}^2}\over{x_1}}+
{{m_{2T}^2}\over{x_2}} \biggr]\Biggl\},
\end{eqnarray}
where  $m_{1T}^2=m_1^2+{\bf q}_T^2$ and $m_{2T}^2=m_2^2+{\bf
q}_T^2$. $m_1$ and $m_2$ are quark or anti-quark masses chosen so
that $m_1\geq m_2$.

It is interesting to note that since the spin of the heavy quark
decouples from the gluon field, the exited states has the same
wave functions as the unexcited ones. The behavior of $\psi_M^2$
is shown in Fig. 1 for the $J/\psi$ state with three different
confinement parameter $\beta$ which are employed to extract the
effect of Fermi motion in this work. Naturally the shape of the
wave function is very sensitive to the values of $\beta$. $x_1$
and $x_2$ both go from 0 to 1 with the constraint of $x_1+x_2=1$.
$q_T$ should cover the necessary range to include the effective
spreading of the wave function.

\begin{figure}
\hskip 1cm \resizebox{.40\textwidth}{!}{%
 \includegraphics{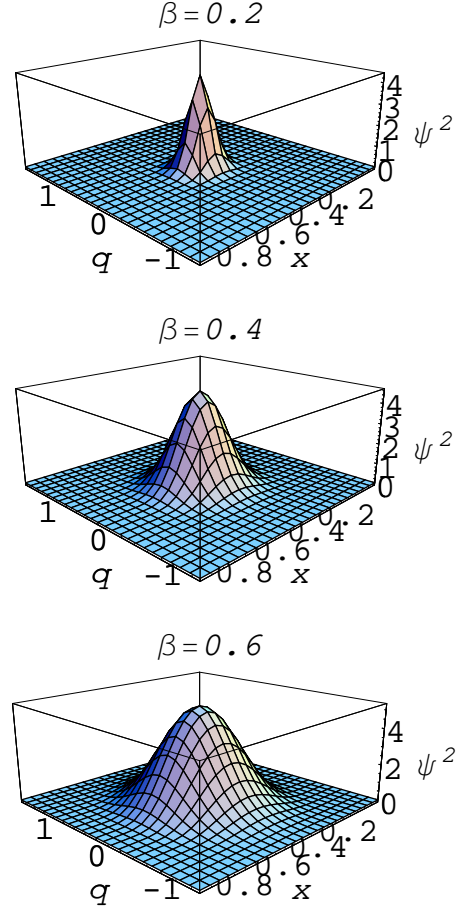}}
\caption[Submanagers]{The behavior of the BHL meson wave function
(11) squared for the $J/\psi$ state with different values of the
confinement parameter. $x$ is the longitudinal energy momentum
ratio and $q$ is the magnitude of the transverse momentum of the
constituents. The scale for $\psi^2$ is arbitrary.} \label{fig:1}
\end{figure}
\section{Kinematics}
\label{sec:3}
 Typical lowest order Feynman diagram for
fragmentation of a $c(b)$ quark into $J/\psi(\Upsilon)$ is shown
in Fig. 2. To evaluate such a diagram consistent with the light
cone wave function for the bound state, we need to choose an
infinite momentum frame in which the initial state heavy quark
possesses transverse momentum of ${\bf p'}_T$. This transverse
momentum is to be carried by the final state $c(b)$ which produces
a jet. The constituents of the meson after creation, move along
the $z$ axis with transverse momentum of ${\bf p}_T=-{\bf k}_T$
($p_T=k_T=q_T$). Therefore the particles' four momenta are
considered as
\begin{eqnarray}
p_{\mu}&=&\bigl[p_{\circ}, \;p_{L}, \;\;{\bf q}_{T}\;\bigr],
\;\;\; k_{\mu}=\bigl[k_{\circ},\; k_{L}, \;-{\bf q}_{T}\;
\bigr],\nonumber\\ p_{\mu}'&=&\bigl[p_{\circ}',\; p'_{L}, \; {\bf
k}_{T}\bigr],\;\; \;\;\; k_{\mu}'=\bigl[k_{\circ}',\; k'_{L}, \;
{\bf k}_{T}\bigr].
\end{eqnarray}
\noindent Note that all particles are moving in the forward
direction and $q_T$'s are chosen in a consistent manner with the
meson wave function. Moreover if we designate the meson four
momentum by $\overline p$, then we may write $p_\mu=x_1\overline
p_\mu$ and $k_\mu=x_2\overline p_\mu$ with the condition that
$x_1+x_2=1$. Therefore, we parameterize the energies of the
particles as
\begin{eqnarray}
 p_{\circ}=x_1 z p'_{\circ}, \; k_{\circ}
= x_2 z p'_{\circ},  \; k'_{\circ} = (1-z) p'_{\circ}, \;
p'_{\circ} =p'_{\circ},
\end{eqnarray}
\noindent where we have used the definition of the fragmentation
parameter as $ z= {E_{ \rm hadron} / {E_{\rm quark}}}$ in an
infinite momentum frame consistent with our kinematics.
\begin{figure}
\hskip -4cm\resizebox{0.85\textwidth}{!}{%
  \includegraphics{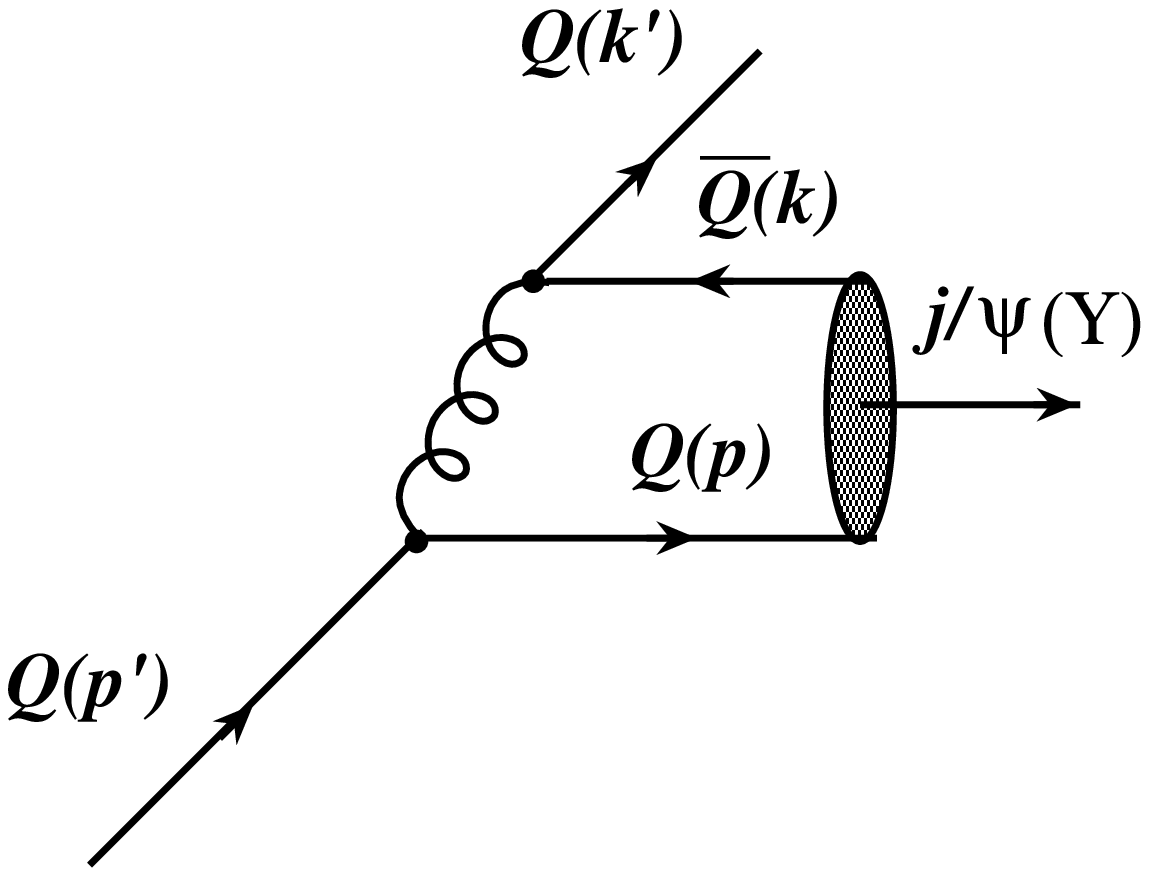}}
\caption[Submanagers]{ The lowest order Feynman diagram for
fragmentation of a heavy quark into a quarkonium state. The four
momenta are labeled.
 }\label{fig:2}
 \end{figure}
\section{The process of fragmentation}
\label{sec:4} The very complex process of a heavy quark
fragmenting into a heavy mesonic bound state is understood and
becomes also calculable if we take into account the following

{$\bullet$} The hard part of the process of a heavy quark
splitting into a heavy meson is calculable in the perturbative QCD
in fragmentation scale [13]. The wave function of the bound state
is convoluted to the hard scattering amplitude in an appropriate
manner.

{$\bullet$} To absorb the soft behavior of the bound state into
hard scattering amplitude a suitable wave function is employed
(our choice (11)).

The hard scattering amplitude for the creation of the bound state
may be put into the following form [14]
\begin{eqnarray}
T_H&=&{{4\pi\alpha_s m_1m_2 MC_F} \over {2\sqrt{2p_\circ p'_\circ
k_\circ k'_\circ}}}\nonumber\\
&& \times{\Gamma \over
{(k+k')^2(p_\circ+k_\circ+k'_\circ-p'_\circ)}}.
\end{eqnarray}
\noindent Here $\alpha_s=g^2/4\pi$ is the strong interaction
coupling constant, $C_F$ is the color factor and $\Gamma$
indicates that part of the amplitude which embeds spinors and
gamma matrices. $(k+k')^2$ is due to the gluon propagator and
$(p_\circ+k_\circ+k'_\circ-p'_\circ)$ is the energy denominator.
In the case of Fig. 2, $\Gamma$ has the following form
\begin{eqnarray}
\Gamma=\overline u(p)\gamma^\mu u(p') \overline u(k') \gamma_\mu
v(k).
\end{eqnarray}
The initial spin average and final state spin sum of the squared
$\Gamma$ takes the following form

\begin{eqnarray}
\frac{1}{2} \sum_s\overline\Gamma \Gamma &=&{1\over{16 m_1^2m_2^2}}\Bigl[(p.k')(p'.k)+(p.k)(p'.k')\nonumber\\
&&+m_2^2(p.p')-m_1^2 (k.k')-2m_1^2m_2^2\Bigr].
\end{eqnarray}

Regarding the kinematics introduced earlier, the dot products in
the above read

\begin{eqnarray}
2p.k'&=&{{1-z}\over{x_1 z}}(m_1^2+q_T^2)+{{x_1
z}\over{1-z}}(m_2^2+k_T^2)\nonumber\\
&&-2{\bf q}_T.{\bf k}_T,
\end{eqnarray}

\begin{eqnarray}
2p'.k={1\over{x_2 z}}(m_2^2+q_T^2)+{{x_2
z}\over{1}}(m_1^2+k_T^2)-2{\bf q}_T.{\bf k}_T,
\end{eqnarray}

\begin{eqnarray}
2p.k={{x_1}\over{x_2}}(m_2^2+q_T^2)+{{x_2
}\over{x_1}}(m_1^2+q_T^2)+2{ q}_T^2,
\end{eqnarray}

\begin{eqnarray}
2p'.k'={{1-z}\over{1}}(m_1^2+k_T^2)+{{1
}\over{1-z}}(m_2^2+k_T^2)-2{ k}_T^2,
\end{eqnarray}

\begin{eqnarray}
2k.k'&=&{{1-z}\over{x_2 z}}(m_2^2+q_T^2)+{{x_2
z}\over{1-z}}(m_2^2+k_T^2)\nonumber\\
&&-2{\bf q}_T.{\bf k}_T,
\end{eqnarray}

\begin{eqnarray}
2p'.p={{1}\over{x_1 z}}(m_1^2+q_T^2)+{{x_1
z}\over{1}}(m_1^2+k_T^2)-2{\bf q}_T.{\bf k}_T.
\end{eqnarray}
Here ${\bf q}_T.{\bf k}_T={ q}_T.{k}_T\cos\theta$ where $\theta$
is the angle between ${\bf q}_T$ and ${\bf k}_T$.

The fragmentation function of a heavy quark fragmentation into a
heavy meson with wave function $\psi$ is obtained from [14]

\begin{eqnarray}
D(z,\mu_\circ) &=&{1\over 2}\sum_s \int d^3p d^3k d^3k' \nonumber\\
&&\times \vert T_H \vert ^2 \vert \psi_M\vert^2 \delta^{(3)}
({p+k+k'-p'}).
\end{eqnarray}
Here $T_H$ is the hard scattering amplitude given by (14) and
$\psi_M$ is the meson wave function (11). Note that average over
initial quark spin and sum over final state particles spin are
assumed. We replace (14) into (23) to obtain

\begin{eqnarray}
D(z,\mu_\circ)&=&{{(4\pi\alpha_s m_1m_2M C_F)^2}\over 8}
\int{{d^3p d^3k d^3k'}\over {p_\circ p'_\circ k_\circ k'_\circ}}\nonumber\\
&&\times{{ {1\over 2}\sum\overline\Gamma
\Gamma\vert\psi_M\vert^2\delta^{(3)} ({p+k+k'-p'}) }\over
{f(z)^2(p_\circ+k_\circ+k'_\circ-p'_\circ)^2}}
\end{eqnarray}
where $f(z)=(k+k')^2$. Now we perform the phase space
integrations. First of all note that

\begin{eqnarray}
\int{{d^3p\delta^{(3)}({p+k+k'-p'})}\over{p'_\circ
(p_\circ+k_\circ+k'_\circ-p'_\circ)^2}}={p'_\circ \over{g(z)^2}},
\end{eqnarray}
where $g(z)=(p+k+k')^2$.When the Fermi motion is ignored,
$\vert\psi_M\vert^2$ is integrated with $d^3 k$ and gives

\begin{eqnarray}
\int {\rm d}^3 k\vert\psi_M\vert^2={{zp'_\circ}\over{16\pi^3}},
\end{eqnarray}
where we have used the normalization condition (3) on the wave
function. Also note that in this case we have

\begin{eqnarray}
\int {\rm d}^3 k'=p'_\circ\int {\rm d}^2 k'_T=2\pi p'_\circ\int
k'_T{\rm d} k'_T .
\end{eqnarray}
Instead of doing this integration, we simply replace $k'_T$ by its
average value. Thus we obtain

\begin{eqnarray}
D(z,\mu_\circ,\beta=0)& =&{{m_1(\alpha_sMC_F)^2\langle
{k'_T}^2\rangle^{1/2}}\over 64 x_1x_2}\nonumber\\
&&\times{{f'}\over {z(1-z)f(z)^2g(z)^2}}
\end{eqnarray}
where $f'$ represents the square bracket in (16). When the Fermi
motion is included, along the same line of calculation we obtain

\begin{eqnarray}
D(z,\mu_\circ,\beta) &=&{{m_1(\pi\alpha_sMC_F)^2\langle
{k'_T}^2\rangle^{1/2}}\over 2} \nonumber\\
&&\times\int{{{ d} q_T { d} x_2 q_T f' \vert\psi_M\vert^2}\over
{x_1x_2 z(1-z)f(z)^2g(z)^2}}.
\end{eqnarray}
Note that the above scheme of the phase space integration allows
one to compare (28) and (29) simply by varying the confinement
parameter $\beta$. Further, note that we have distinguished the
quark and anti-quark masses such that (28) and (29) may be used in
the case of other meson states such as $D$, $B$ and $B_c$.

The explicit form of the fragmentation functions (28) and (29) in
general are rather lengthy. However in the case of quarkonia they
cast into rather brief forms. They appear in the appendix.

\section{Results}
\label{sec:5} We have shown that the wave function due to BHL may
be employed to represent the soft behavior of the heavy meson
bound state. In infinite momentum frame we have set up a
kinematics which embed the Fermi motion of the constituents in
both longitudinal and transverse direction through this wave
function whose width is controlled by the confinement parameter
$\beta$. Comparison of (28) and (29) demonstrates the effect in an
illustrative manner. This comparison is shown in Figure 3 in the
case of the $J/\psi$ and $\Upsilon$ states. The two functions
coincide at sufficiently small values of $\beta$. As we increase
$\beta$, the peak due to (28) rises and $\langle z \rangle$ moves
to higher values. This behavior is consistent with the physics of
the effect. In fact inclusion of Fermi motion in the fragmentation
of a particular state means fragmentation of a state with higher
energy and momentum which naturally ends up with higher values of
both the fragmentation probability and average fragmentation
parameter. Note the ranges of $\beta$ selected in the two diagrams
in this figure. They are the same as those used in Figure 1. To
find the size of such changes, we employ the fragmentation
probability defined by

\begin{eqnarray}
P(\beta)=\int D(z,\mu_\circ,\beta) dz,
\end{eqnarray}
and the average fragmentation parameter

\begin{eqnarray}
\langle z \rangle(\beta)={{\int  z D(z,\mu_\circ,\beta)dz }\over
{\int D(z,\mu_\circ,\beta) dz}}.
\end{eqnarray}
We plot $P(\beta)/P(0)$ versus $\beta$ in Figure 4. Such
evaluations reveal that the Fermi motion of the constituents of
$J/\psi$ and $\Upsilon$ states improve the fragmentation
probabilities of these states at least up to 14 percent. We have
also plotted $\langle z \rangle(\beta)$ against $\beta$ in Figure
5.
\begin{figure}
\hskip 1.cm\resizebox{0.38\textwidth}{!}{%
  \includegraphics{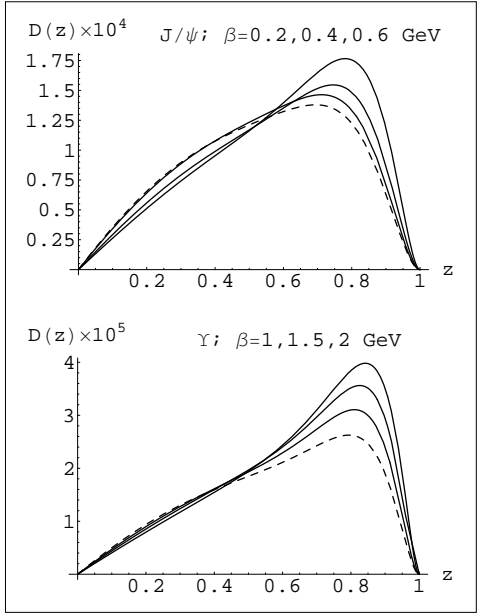}}
\caption[Submanagers]{The affect of $J/\psi$ and $\Upsilon$ wave
function on their fragmentation. While the dashed curve represent
(28), the solid curves show the behavior of (29) with increasing
values of $\beta$. The two curves coincide exactly for
sufficiently small values of confinement parameter. }\label{fig:3}
\end{figure}

\begin{figure}
\hskip 1.cm\resizebox{0.38\textwidth}{!}{%
  \includegraphics{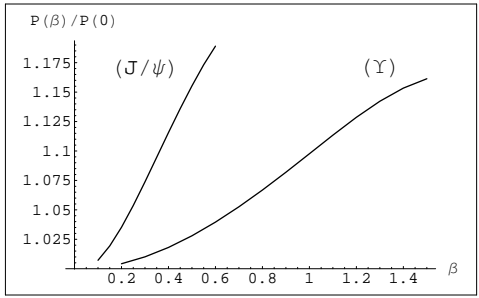}}
\caption[Submanagers]{Rate at which the fragmentation
probabilities for $J/\psi$ and $\Upsilon$ increase with increasing
confinement parameter. Note the ranges of $\beta$ selected for the
two cases. }\label{fig:4}
\end{figure}
\begin{figure}
\hskip 1.cm\resizebox{0.38\textwidth}{!}{%
  \includegraphics{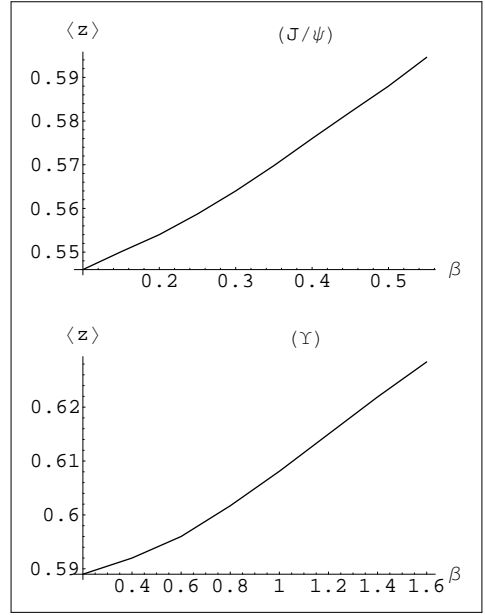}}
\caption[Submanagers]{Average fragmentation parameter $\langle
z\rangle$ versus  confinement parameter $\beta$ for $J/\psi$ and
$\Upsilon$.  }\label{fig:5}
\end{figure}

This investigation shows that the Fermi motion within the heavy
meson states such as $J/\psi$ and $\Upsilon$ is indeed important
and could not be ignored. This of course will be reflected on the
cross section and event rate of these states at the colliders
where they are expected to be produced.

Concerning the $J/\psi$ production our results could be accounted
for as a modification to the color singlet mechanism where
contribution due to various channels specially the color octet
mechanism are relevant.
{}
\vskip .5cm \noindent{\bf\large Appendix} \vskip .5 cm

\noindent The explicit form of (28) and (29) for quarkonia cast
into the following

\begin{eqnarray}
D(z,\mu_\circ,\beta=0)& =&{{\alpha_s^2 C_F^2\langle
{k'_T}^2\rangle^{1/2}}\over 16 m {\cal F}}\nonumber\\
&&\times\Bigl\{ z(1-z)^2\bigl[\xi^2z^4+2\xi z^2
(4-4z+5z^2)\nonumber\\ &&+
(16-32z+24z^2-8z^3+9z^4)\bigr]\Bigr\}\nonumber,
\end{eqnarray}
and

\begin{eqnarray}
D(z,\mu_\circ,\beta) &=&{{\pi^2\alpha_s^2 C_F^2\langle
{k'_T}^2\rangle^{1/2}}\over {2 m }}\nonumber\\
&&\times\int \frac{dq dx  q x^2 z (1-z)^2 |\psi_M|^2}{{\cal
G}}{{\cal H}},\nonumber
\end{eqnarray}
where ${\cal F}$, ${\cal G}$ and ${\cal H}$ read as

\begin{eqnarray}
{\cal F} =\bigl[\xi^2z^4-(z-2)^2(3z-4)+\xi
z^2(8-7z+z^2)\bigr]^2,\nonumber
\end{eqnarray}

\begin{eqnarray}
{\cal G}&=&\Bigl\{\bigl[\eta (1-z)^2+\xi x^2z^2+(1-(1-x)z)^2\bigr]\nonumber\\
&&\times\bigl[\eta(-1+z)+\xi
(-1+x)xz^2-1+(1-x+x^2)z)\bigr]\Bigr\}^2,\nonumber
\end{eqnarray}
\noindent and
\begin{eqnarray}
{\cal H}&=&\Bigl\{1-4(1-x)z+2(4-10x+7x^2)z^2+4(-1+4x\nonumber\\
&&-5x^2+x^3)z^3+(1-4x+8x^2-4x^3+x^4)z^4+\eta\xi z^2\nonumber\\
&&\times\bigl[1-2x+z^2+x^2(2-2z+z^2)\bigr]+ \eta\bigl(2+(4x-6)z\nonumber\\
&&+(9-8x+2x^2)z^2-2(2-x+x^2)z^3+(1+x^2)z^4\bigr)\nonumber\\
&&+\xi z^2\bigl(1+2x^3(2-3z)z+z^2+2x^4z^2+x(-2+2z\nonumber\\
&&-4z^2)+x^2(2-8z+9z^2)\bigr)+\eta^2(1-z)^2\nonumber\\
&&+\xi^2(1-x)^2x^2z^4\Bigr\}.\nonumber
\end{eqnarray}
Here $\xi=\langle {k'_T}^2\rangle/m^2$ and  $\eta= {q}^2/m^2$.
$A_M$ is calculated using the normalisation condition of the wave
function, e.g. equatrion (3).


\begin{thebibliography}{}
\bibitem{RefJ} R.Baier and R. R\"{u}ckl, Phys. Lett. B {\textbf 102},
(1981) 364; C.H. Chang, Nucl. Phys. B {\textbf 172}, (1980) 425;
C.E. Carlson and R. Suaya, Phys. Rev. D {\textbf 14}, (1976) 3115;
M.B. Einhorn and S.D. Ellis, Phys. Rev. D {\textbf 12} (1975)
2007.
\bibitem{RefJ} R. Baier and R. R\"{u}ckl, Z. Phys. C {\textbf 29},
(1983) 251; E.L. Berger and D. Jones, Phys. Rev. D {\textbf 23},
(1981)1521;  S.D. Ellis, M. Einhorn and C. Quigg, Phys. Rev. Lett.
{\textbf 36} (1976) 1263.

\bibitem{RefJ} G.T. Bodwin, E. Braaten and G. P. Lepage, Phys. Rev.
D {\textbf 51} (1995) 1125; E. Braaten M.A. Doncheski, S. Fleming
and M. Mangano, phys. lett. B {\textbf 333}, (1994) 548; G.T.
Bodwin, E. Braaten and G.P. Lepage, Phys. Rev. D {\textbf 46}
(1992) R1914.

\bibitem{RefJ} E. Braaten, S. Fleming and T.C. Yuan, Ann. Rev.
Nuc. Part. Sci. {\textbf 46}, (1996) 197.

\bibitem{RefJ} P. Cho and A.K. Leibovich, Phys. Rev. D {\textbf 53} (1996)
6203.

\bibitem{RefJ} E. Braaten {\it et al}, Phys. Rev. D {\textbf 48},
(1993) 5049 ;E. Braaten {\it et al}, Phys. Rev. D {\textbf 51},
(1995) 4819; M.A. Gomshi Nobary and T. O'sati, Mod. Phys. Lett. A
{\textbf 15}, (2000) 455.

\bibitem{RefJ}G. Altarelli and G. Parisi, Phys. Lett. B {\textbf
71}, (1977) 298.

\bibitem{RefJ} Leonid Frankfort and Werner Koeff, Phys. Rev. D {\textbf
54}, (1996) 3194; Leonid Frankfort, Werner Koeff and Mark
Strikman, Phys. Rev. D {\textbf 57}, (1998) 512

\bibitem{RefJ}S.J. Brodsky, T. Huang and G.P. Legage,
in {\it Particles and Fields}, Proceedings of the Banff
Summer Institute, Banff, Alberta, 1981,edited by A.Z. Capri and
A.N. Kamal (Plenum, New York, 1983), p. 143; Xin-heng Guo and Tao
Huang, Phys. Rev. D {\textbf 43}, (1993) 2931 ; F. Schlumpf, Phys.
Rev. D {\textbf 50}, (1994) 6895

\bibitem{RefJ} Xin-heng Guo and Tao Huang, Phys. Rev. D {\textbf
43}, (1991) 2931

\bibitem{RefJ}Tao Huang {\it et al}, Phys. Rev. D {\textbf 49},
(1994) 1490.

\bibitem{RefJ} See, e.g. Elementary Particle Theory Group, Acta
Phys. Sin. {\textbf 25}, (1976) 415; N. Isgur, in {\it New Aspects
of Subnuclear Physics}, edited by A. Zichichi (Plenum, New York,
(1980), p. 107.

\bibitem{RefJ} See for example E. Braaten {\it et al} in Ref. [6]

\bibitem{RefJ} M. Suzuki, Phys. Rev. D {\textbf 33}, (1986) 676.


\end{thebibliography}
\end{document}